\documentclass[sigconf]{acmart}

\usepackage{booktabs} 
\usepackage{multirow}
\usepackage{graphicx}
\usepackage{caption}
\usepackage{subcaption}

\setcopyright{none}

\begin{document}
\title{Discovering Scholarly Orphans Using ORCID}

\author{Martin Klein}
\affiliation{Los Alamos National Laboratory}
\affiliation{Los Alamos, NM, USA}
\affiliation{\url{http://orcid.org/0000-0003-0130-2097}}
\email{mklein@lanl.gov}
\author{Herbert Van de Sompel}
\affiliation{Los Alamos National Laboratory}
\affiliation{Los Alamos, NM, USA}
\affiliation{\url{http://orcid.org/0000-0002-0715-6126}}
\email{herbertv@lanl.gov}

%
%
\begin{abstract}
Archival efforts such as (C)LOCKSS and Portico are in place to ensure the longevity of 
traditional scholarly resources like journal articles. 
At the same time, researchers are depositing a broad variety of other scholarly artifacts 
into emerging online portals that are designed to support web-based scholarship. 
These web-native scholarly objects are largely neglected by current archival practices
and hence they become scholarly orphans.
We therefore argue for a novel paradigm that is tailored towards archiving these scholarly
orphans. 
We are investigating the feasibility of using Open Researcher and Contributor ID (ORCID) 
as a supporting infrastructure for the process of discovery of web identities and 
scholarly orphans for active researchers. We analyze ORCID in terms of 
coverage of researchers, subjects, and location and assess the richness of its profiles 
in terms of web identities and scholarly artifacts.
We find that ORCID currently lacks in all considered aspects and hence can only be considered
in conjunction with other discovery sources. However, ORCID is growing fast so there is potential
that it could achieve a satisfactory level of coverage and richness in the near future.
\end{abstract}
%
%
%
%
%
%
\maketitle
\section{Introduction} \label{sec:intro}
Over the past two decades, research communication has transitioned from a paper-based endeavor 
to a web-based digital enterprise. More recently, the research process itself has started to 
evolve from being a largely hidden activity to one that becomes plainly visible on the global 
network. 
To support researchers in this process, a wide variety of online portals have emerged which largely
exist outside the established scholarly publishing system. These portals can be dedicated to
scholarship, such as \url{experiment.org}, or general purpose, such as \url{SlideShare.net}.
The ``$101$ Innovations in Scholarly Communication'' 
project\footnote{\url{https://101innovations.wordpress.com/}} provides a first of a kind overview 
of such platforms. 
The large number of readily available web portals promts some to even argue that there are too many
of them, leading to decision fatigue \cite{tattersall2017}.
Regardless, the potential of increased productivity and global exposure attracts researchers and so 
they happily deposit scholarly artifacts there. 

However, history has shown that even popular web platforms can disappear without a trace. To
make matters worse, they rarely provide any explicit archival guarantees; many times quite the opposite. 
Whereas initiatives such as LOCKSS\footnote{\url{https://www.lockss.org/}} and 
Portico\footnote{\url{http://www.portico.org/digital-preservation/}} have emerged to make sure 
that the output of the established scholarly publishing system gets archived, to the best of our
knowledge, no comparable efforts exist for scholarly artifacts deposited in these online platforms. 
We are therefore motivated to explore how these scholarly artifacts deposited in online portals 
could be archived. 
\subsection{Current Archival Paradigm}
To a large extent, the paradigm that underlies current approaches to capture and
archive web-based scholarly resources has its origin in the paper-based era. It can
be characterized as a back-office procedure in which the owner of a scholarly object
decides when to hand over a finalized and atomic object to a custodian that will take
care of its long-term preservation. The transfers by a publisher of its journals to
Portico and the upload of an article by its author into an Institutional Repository are
examples of such procedures. 
However, we see several signs indicating that this paradigm's capture approach is failing 
even for journal articles, the most traditional of scholarly resources. David Rosenthal, 
amongst others, has reported that a significant portion of journal articles does not make it into 
an archive and several reasons can be attributed to that \cite{dshr2013,dshr2015}.
He observes, for example, an apparent focus on articles that are technically not too complex to 
capture and those published by large publishers.
To make matters worse, this traditional paradigm insufficiently accounts for the fact that journal 
articles no longer exhibit inherent fixity but rather are ``living things'' with versions. It also
does not incorporate attempts to capture web content that is directly related to journal articles 
i.e., web resources linked from these articles \cite{klein:one_in_five, jones:content_drift}. 
The reason for this failure is probably the fact that journal articles are largely still regarded 
as static atomic objects despite the overwhelming evidence that they have become dynamic and firmly 
embedded in the web.
\subsection{Exploring a Novel Paradigm}
We postulate that a paradigm that fails for the most traditional scholarly outputs is highly likely 
to fail when novel, web-native scholarly objects used in research communication and the research 
process are at stake. Such objects include all sorts of scholarly artifacts deposited in web portals
such as slide decks, videos, simulations, software, workflows, and ontologies. 
Since these web-native scholarly objects are largely neglected by the current archival 
paradigm \cite{dshr2015}, we refer to them as \textbf{scholarly orphans}. 
They also have dramatically different characteristics than traditional articles or monographs in 
that they are compound (aggregations of related resources), dynamic (versioning), interdependent, 
distributed across the web \cite{sompel:interop,bechhofer:research_objects}, and created at another 
scale altogether.

We therefore argue for a new archival paradigm. We envision an archival paradigm inspired by web 
archiving concepts that is web-centric to be able to cope with the scale of the problem, both in 
terms of the number of platforms and the number of artifacts involved.
Because the artifacts are often times created by researchers affiliated with an institution, 
we assume that these institutions are interested in collecting the artifacts. Therefore,
and for the sake of efficiency and scale, we explore a new archival paradigm built around highly 
automated web-scale processes operated on behalf of a scholarly institution.
\subsection{Outline of a Novel Archival Paradigm}
\begin{figure}[t!]
\includegraphics[scale=0.5]{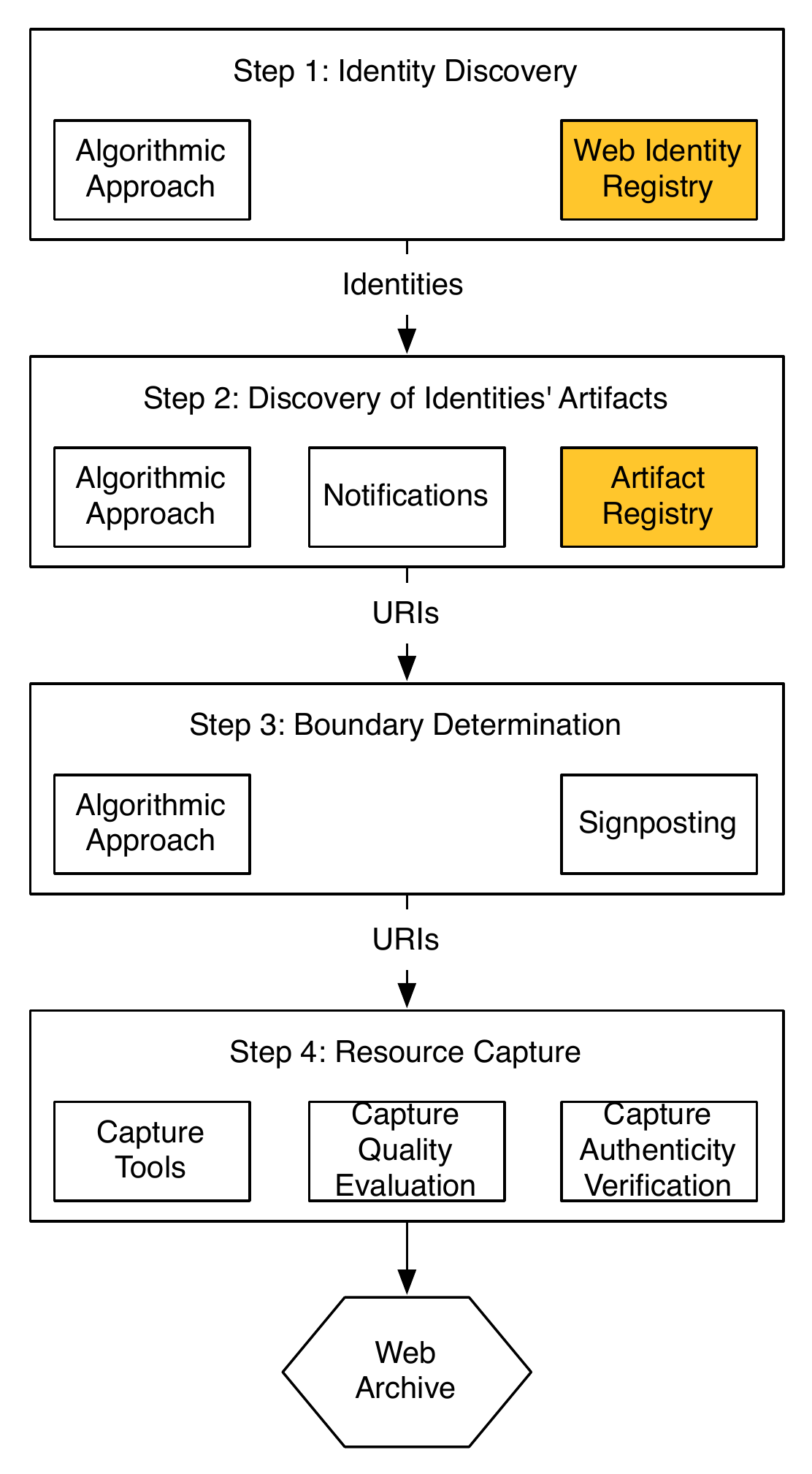}
\caption{High-level conceptual overview of the envisioned process to discover, scope, and capture scholarly orphans}
\label{fig:paradigm_concept}
\end{figure}

A conceptual view of the high-level processes in our paradigm under exploration is depicted in 
Figure \ref{fig:paradigm_concept}.
\begin{enumerate}
\item The first step is to discover the web identities of institutional scholars in various online portals
such as SlideShare handles, FigShare names, etc. 
This can either be achieved with an algorithmic approach, for example, by using web discovery on the basis 
of metadata about the scholar \cite{powell:egosystem} or by means of registries that list researcher
profiles such as ORCID.
\item The second step, which builds on the web identities discovered in step 1, is to discover actual 
artifacts created or contributed to by the scholar. The discovery of the artifacts on the basis of 
those web identities largely depends on the functionality of the portal. One option is to subscribe to the
portal's notification service that, if available, sends messages whenever new objects are created. An
alternative is to recurrently visit a registry e.g., a list of artifacts that indexes the scholar's artifacts 
deposited in the portal. If neither of these options are available, an algorithmic approach could also 
be deployed here.
\item Several resources, each with their distinct URI, may pertain to any given artifact. As such, in 
order to capture the entire artifact, its web boundary - the list of all URIs that pertain to the 
artifact - must be determined. This can either be done in an algorithmic manner, which requires 
extensive portal-specific heuristics \cite{sompel:infrastructure} or by means of information explicitly 
exposed by the portals in manners proposed by, for example, Signposting\footnote{\url{http://signposting.org/}} 
and OAI-ORE\footnote{\url{http://www.openarchives.org/ore/1.0/datamodel}} \cite{lagoze:ore}. 

\item The final step in the process is the capture of discovered artifacts, that is, capture all 
URIs that are within the web boundary of the artifact. A variety of tools have 
emerged from the web archiving community that could be used for the capture such as 
Heritrix\footnote{\url{https://webarchive.jira.com/wiki/display/Heritrix}}, 
Brozzler\footnote{\url{https://github.com/internetarchive/brozzler}}, 
Webrecorder\footnote{\url{https://webrecorder.io/}}, and iCrawl \cite{gossen:icrawl}.
that can be deployed here. To accommodate concerns regarding the quality and trustworthiness of 
captures, this step can also include a capture quality evaluation and a capture authenticity 
verification.
\end{enumerate}
\subsection{ORCID}
A detailed analysis of all of the components of these processes outlined above is beyond the 
scope of this paper. The focus of this paper is on determining whether Open Researcher and 
Contributor ID (ORCID), a rapidly growing database of scholarly web identities (ORCIDs) and 
associated profiles can play a role in the archival paradigm that we explore and that is 
depicted in Figure \ref{fig:paradigm_concept}.

The ORCID database has become increasingly popular since its inception in $2012$. At the time 
of writing it registers just over three million profiles. Its core motivation was to solve the 
issue of name disambiguation and provide a platform for the unique identification of contributors 
to scholarly work \cite{haak:orcid}. Scholars are motivated to create, populate, and maintain 
their profile to advertise their accomplishments and gain credit for them.
Publishers and funding agencies are also recognizing the merit of ORCIDs and have begun to mandate 
their inclusion in papers and project proposals.
However, in the greater picture of scholarly communication the ORCID platform has the potential 
to emerge as crucial infrastructure to unambiguously bind scholars to their work. In addition, 
implementations emerge that allow researchers to authenticate against scholarly 
portals with their ORCID and use the same identity in many different platforms. Consequently, 
opportunities arise to bind a researcher's scholarly web identity to other web identities.

We believe that the ORCID platform has enormous potential to play a core role as a web identity 
registry in step 1 and as an artifact registry in step 2 of the archival paradigm (top two boxes 
in Figure \ref{fig:paradigm_concept}) that we explore. 
However, in order for ORCID to be able to play such a role, the platform must have substantial 
coverage of active scholars and rich scholar profiles.
We investigate the suitability of ORCID for this purpose and to make this assessment we ask the 
following research questions:
\begin{enumerate}
\item Does the ORCID platform represent the broadest possible coverage of researchers, in absolute 
numbers, coverage of subjects, and coverage per geographical area? ($RQ1$)
\item Are ORCID profiles rich with information about the scholar that is useful for our cause 
as well as web identities and artifacts? ($RQ2$)
\end{enumerate}
Addressing these two questions ($RQ1$ and $RQ2$) combined with offering insight into the evolution of 
ORCID adoption and ORCID profiles over time is the main contribution of this paper. We conduct a 
study to evaluate ORCID records over time to assess whether trends support our intuition that ORCIDs 
could be leveraged in steps 1 and 2 of our archival paradigm.
%
%
%
\section{Related Work}
Given the novel and exploratory nature of this work, to the best of our knowledge, there are no 
comparable efforts in this realm that are addressing the same issues.
However, web-centric archiving of scholarly resources is not a novel concept. The Lots of Copies
Keep Stuff Safe (LOCKSS) program 
is built on open source
peer-to-peer technology \cite{rosenthal:lockss} to focus on preserving scholarly content for 
long-term access.
The recent work by Van de Sompel, Rosenthal and Nelson \cite{sompel:infrastructure} outlines a
multitude of problems with this regard. For example, the fact that e-journal preservation systems
have to spend a lot of time and effort on developing crawlers that grab articles from publishers' 
websites. This is a time-consuming and hence expensive endeavor that requires a lot of expert 
knowledge about a publisher's website structure, especially when dealing with the long tail 
of smaller publishers.
The LOCKSS system is relevant to our paradigm but not directly comparable since we are targeting
scholarly orphans, artifacts that are neglected by existing archival approaches.

The EgoSystem \cite{powell:egosystem} developed at the Los Alamos National Laboratory was designed
to discover web identities of the lab's postdoctoral students. It used basic information about the
student such as name, degree-awarding institution and the student's field of study as the seed to 
search for web identities via the Yahoo! search engine and in a pre-defined list of social and 
academic web portals. In its initial phase, EgoSystem targeted web identities within Microsoft 
Academic, LinkedIn, Twitter, and SlideShare but also searched for personal homepages and Wikipedia 
articles. Not only did EgoSystem sucessfully return a list of web identities, it also kept a record 
of search results and learned additional associations with every new query.
Northern and Nelson \cite{northern:unsupervised} developed an unsupervised approach to discover web
identities on social media sites. The discovery phase was based on queries to search engines and to
social media sites directly with the name of an individual as well as with variations of the name. 
The process also included an disambiguation step that was based on comparing key features extracted
from discovered candidate profiles. 
Both systems are related to our approach as they offer approaches for the algorithmic discovery of
web identities, even if the motivation to do so was different from ours. It is worth noting that
services essential to the operation of both systems are no longer available. For example, the 
Yahoo! Search API as well as the Microsoft Academic API have been discontinued.
%
%
%
%
\section{Experiment Setup}
The ORCID organization publishes high-level statistics and updates them on a regular 
basis\footnote{\url{https://orcid.org/statistics}}. Amongst the statistics are the total 
number of ORCIDs, the number of ORCIDs with at least one ``work'' (reference to a publication,
dataset, patent, or other research output), employment as well as education activities.
The ORCID organization has been providing data dumps of all records and all publicly available 
information within these records once a year since $2013$. We were therefore able to download all 
available datasets of ORCID records from $2013$ \cite{orcid2013}, $2014$ \cite{orcid2014}, 
$2015$ \cite{orcid2015}, and $2016$ \cite{orcid2016}.
Table \ref{tab:orcid_data} summarizes the size of the obtained ORCID datasets and the number of ORCID records 
they contain. Each dataset represents a snapshot of ORCID records at a particular point in time. For example,
the $2016$ dataset contains all records as of October 1st $2016$.
The datasets contain two serializations for each ORCID record, one in XML and one in JSON format, and
we chose to work from the JSON files.
\begin{table}[t!]
\caption{Public ORCID datasets}
\begin{tabular}{|c|c|c|} \hline
\textbf{Year} & \textbf{File Size in GB} & \textbf{ORCID Records} \\ \hline \hline
$2013$ & $0.62$ & $361,209$ \\ \hline
$2014$ & $2.2$ & $910,470$ \\ \hline
$2015$ & $5.9$ & $1,588,199$ \\ \hline
$2016$ & $11.0$ & $2,528,933$ \\ \hline \hline
$\sum$ & $19.72$ & $5,388,818$ \\ \hline
\end{tabular}
\label{tab:orcid_data}
\end{table}
%
%
%
%
%
%
%
%
\subsection{Data Preparation and Enrichment for ORCID Coverage ($RQ1$)}
To approach $RQ1$ we investigate ORCID coverage in terms of absolute number of researchers,
in terms of subjects, and in terms of geographical coverage. To do so, we extract particular
data from the ORCID profiles. 
All we need to assess the coverage of number of reseachers is the total number of ORCID records
in a dataset. This data is available from Table \ref{tab:orcid_data}.
In order to evaluate the geographical coverage of ORCID records, we extract the most recent 
affiliation information from all profiles. This data not only comes with the name of the institution
but also with its location. We are therefore able to map the distribution of locations (by a country
granularity) from ORCID profiles.
To determine the subject coverage, however, a more elaborate data preparation process is needed.
We first extract all available information about scholars' works, in particular, the name of the 
author(s), the title, the publication year, and, if provided, the works's DOI.
Since the works records in ORCID profiles do not contain subject information, we need to acquire
this information from another source. 

The CrossRef Metadata Search 
API\footnote{\url{https://github.com/CrossRef/rest-api-doc/blob/master/rest_api.md}} returns 
metadata about DOI-identified scholarly objects such as title, author, publisher and license 
information, etc.  In addition, it provides a set of subject terms describing the work and its field 
of study.  These subject terms are provided by the publisher and therefore not all of them 
neccessarily adhere to the same ontology. However, it is not unreasonable to assume that individual 
publishers use the same set of subject terms for all their papers. For example, two papers in the 
area of high-energy physics that are published by the same publisher are very likely both be assigned 
the subject ``Physical Science''.
We utilize this service and query all extracted DOIs against the API and extract the returned 
subject terms for each work.
To unify the results, we are in need of a standardized set of subjects.
Fortunately, the Classification of Instructional Programs (CIP) published by the Institute of 
Education Sciences' National Center for Education 
Statistics\footnote{\url{https://nces.ed.gov/ipeds/cipcode/Default.aspx}} offers just that. 
The CIP provides a taxonomy that is made up of $47$ high-level subjects (each having multiple
finger-granularity subjects) that maps the most common fields of study. 
We can therefore match our subject terms obtained from CrossRef against the CIP subjects.
The matching is based on simple word comparison after minor pre-processing such as transforming all
strings to upper case and ignoring trailing quantifiers such as ``Other'' and ``General''. For example,
we transformed the CIP subject ``Agricultural Business and Management, Other'' into simply
``Agricultural Business and Management''. To decrease the granularity of subjects, we bin all matches of
a lower level subject into the highest level subject. For example, if a DOI matches the lower level
subject ``Agricultural Business and Management'' (CIP code $01.0199$), it is binned into the highest 
level subject ``Agriculture, Agriculture Operations, and Related Sciences'' (CIP code $01$).
\subsection{Data Preparation and Enrichment for Richness of Profiles ($RQ2$)} \label{sec:data_prep_rq2}
$RQ2$ aims at investigating the richness of ORCID profiles in terms of web identities, artifacts of 
interest to our archiving paradigm, and other information they contain about the scholar.
The data preparation processes here are fairly straight forward. We extract data on web identities 
as well as other information about the scholar (for example, given and family name, affiliation)
from the metadata section of each profile. To assess the suitability of the web identities for our
purpose, we extract and analyze their associated labels.
We further obtain the type information for all artifacts in order to evaluate whether they are in 
scope for our new archival paradigm.

It is worth noting that the information in an ORCID profile can be subject to access restrictions if 
the owner choses to establish them. However, after an email exchange with the ORCID customer support, 
we can confirm that the majority of data we are interested in is publicly accessible. For 
example, $88.6\%$ of works, $96.2\%$ of names, and $87.0\%$ of affiliations do not have any access 
restrictions.
\section{ORCID Coverage ($RQ1$)}
To address $RQ1$, we investigate to what extend ORCID covers a broad spectrum of reseachers, 
subjects, and geographical locations. 
\subsection{Coverage of ORCID in Absolute Numbers} \label{sec:cov_abs_numbers}
Our first coverage-related investigation is on the raw numbers of ORCID records and how that compares
to the total number of researchers worldwide. 
The latest UNESCO Science Report published in $2015$ \cite{unesco2015} states that in $2013$
there were $7,758,900$ researchers worldwide. As shown in Table \ref{tab:orcid_data}, even the largest
ORCID dataset from $2016$ holds $2,528,933$ profiles, only about one third of the total number of 
researchers.
The UNESCO report also provides the total number of researchers in the U.S. only. For the year $2013$
this number is at $1,265,100$. In comparison, by extracting metadata from the $2016$ dataset, we find 
a total of only $112,577$ ORCID profiles that list their most recent affiliation as located in the 
U.S., which equals $8.9\%$. 
Neither the comparison worldwide nor the one specific to the U.S. indicates that ORCID has a 
representative coverage of researchers, provided in absolute numbers.

It is worth noting though that the number of ORCID records is growing at a faster pace than the number 
of researchers.
As shown in Table \ref{tab:orcid_data}, the increase of profiles initally is very steep 
with more than $2.5$ times as many records in $2014$ as in $2013$. The increase of $74\%$ in $2015$ and 
$59\%$ in $2016$ is still significant compared to the respective previous years. 
Worldwide, the growth in number of researchers has been betweem $5\%$ and $7\%$ since $2007$.
If these trends continue, there is a potential for ORCID to achieve full coverage by $2020$.
%
%
%
%
%
%
\subsection{Subjects Covered by ORCID}
Our second part of the ORCID coverage analysis focuses on the coverage of research subjects.
We obtain data about the number of recipients of doctorate degrees as well as the number of 
scientific publications as a proxy to assess subject coverage.

First, in order to assess the subject distribution of ORCIDs, we need to compute the subjects covered 
by each individual scholar with an ORCID identity on the basis of the CIP terms obtained via her 
DOI-identified artifacts, as described above. 
If an ORCID only has one DOI associated with it that matches against one CIP term only,
this ORCID provides the score of $1$ to the matched subject.
However, it is entirely possible that one publication falls into multiple areas of study, is 
associated with multiple subject terms from the publisher, and hence is matched against more than 
one CIP subject. In this case we distribute the subject score for that DOI accordingly. For example, 
if a DOI matches two the subjects ``Agriculture, Agriculture Operations, and Related Sciences'' 
and ``Education'' (CIP code $13$), both of these subjects get a score of $0.5$. The sum of 
the matches per DOI is always $1$, so if a DOI matches three subjects, each receives a score 
of $1/3$. We aggregate all scores per subject and rank them in decreasing order of their scores.

\begin{figure}[t!]
\includegraphics[scale=0.4]{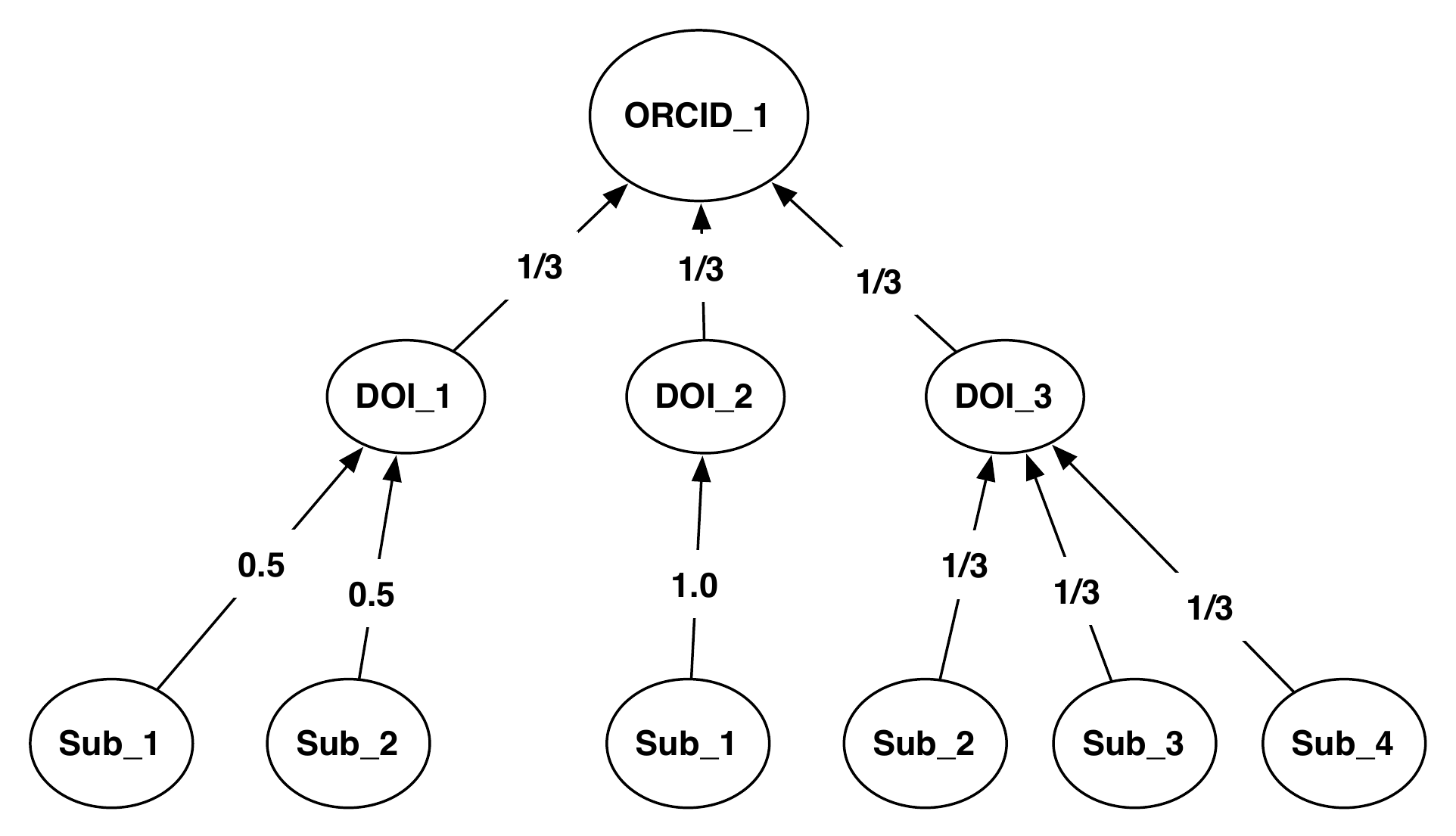}
\caption{Subject score aggregation per ORCID record}
\label{fig:orcid_doi_subject_tree}
\end{figure}

To assess the distribution of subjects for all DOI-identified artifacts contributed by a single 
researcher, we aggregate the individual DOI scores per ORCID. 
Figure \ref{fig:orcid_doi_subject_tree} showcases an example where an ORCID 
record ($ORCID_1$) has three DOI references, $DOI_1$, $DOI_2$, and $DOI_3$.
$DOI_1$ matches two subjects ($Sub_1$ and $Sub_2$) so each of them score $0.5$ for the DOI. However,
on the level on ORCID records, since $ORCID_1$ has three DOIs, these scores are weighted with a 
factor of $1/3$. $DOI_2$ only matches $Sub_1$ and so its score is $1.0$ before being weighted on 
the ORCID record level. $DOI_3$ matches $Sub_2$, $Sub_3$, and $Sub_4$ and hence each of the 
subjects get a score of $1/3$ before individually weighted on the ORCID record level.
Table \ref{tab:orcid_score} summarizes the computation and results for each subject on this
example ORCID. Similar to the level of individual DOIs, the sum of the subject scores per ORCID 
is always $1$.
\begin{table}[t!]
\caption{Subject scores for example ORCID record $ORCID_1$}
\begin{tabular}{|c|c|c|} \hline
\textbf{Subject} & \textbf{Computation} & \textbf{Result} \\ \hline \hline
$Sub_1$ & $(0.5\times1/3)+(1.0\times1/3)$ & $1/2$\\ \hline
$Sub_2$ & $(0.5\times1/3)+(1/3\times1/3)$ & $5/18$ \\ \hline
$Sub_3$ & $(1/3\times1/3)$ & $1/9$ \\ \hline
$Sub_4$ & $(1/3\times1/3)$ & $1/9$ \\ \hline
\end{tabular}
\label{tab:orcid_score}
\end{table}

To the best of our knowledge there is no comprehensive list with numbers of researchers by area of study. 
We therefore use the numbers of awarded Ph.D. degrees in the U.S. as an estimation for the distribution of
reseachers' disciplines. The National Science Foundation (NSF) regularly publishes a report on doctorate 
recipients from U.S. universities\footnote{\url{https://www.nsf.gov/statistics/2017/nsf17306/datatables/tab-12.htm}}
from which we extract the $2015$ data.
The report classifies all recipients' disciplines into subjects that are very similar to the CIP subjects 
we used and hence can easily be compared. 
We take the relative numbers of recipients by subject and compare this data to the relative score distribution 
of subjects derived from publications in ORCID records.

We further obtain the total numbers of scientific publications in the U.S. in $2014$ from the same
UNESCO Science Report \cite{unesco2015} mentioned earlier.
Similar to the NSF data described above, this report also classifies all publications into subjects that 
are very similar to the CIP subjects we used. We extract the numbers from the UNESCO report and compute 
the relative numbers of publications by subject. Note that the UNESCO report does not maintain specific 
data for the fields of ``Education'' and ``Humanities and Arts''. It is likely that publications from 
these areas are binned into the generic ``Other'' category and hence prohibits a comparison for them.

\begin{figure}[t!]
\centering
\includegraphics[scale=0.35]{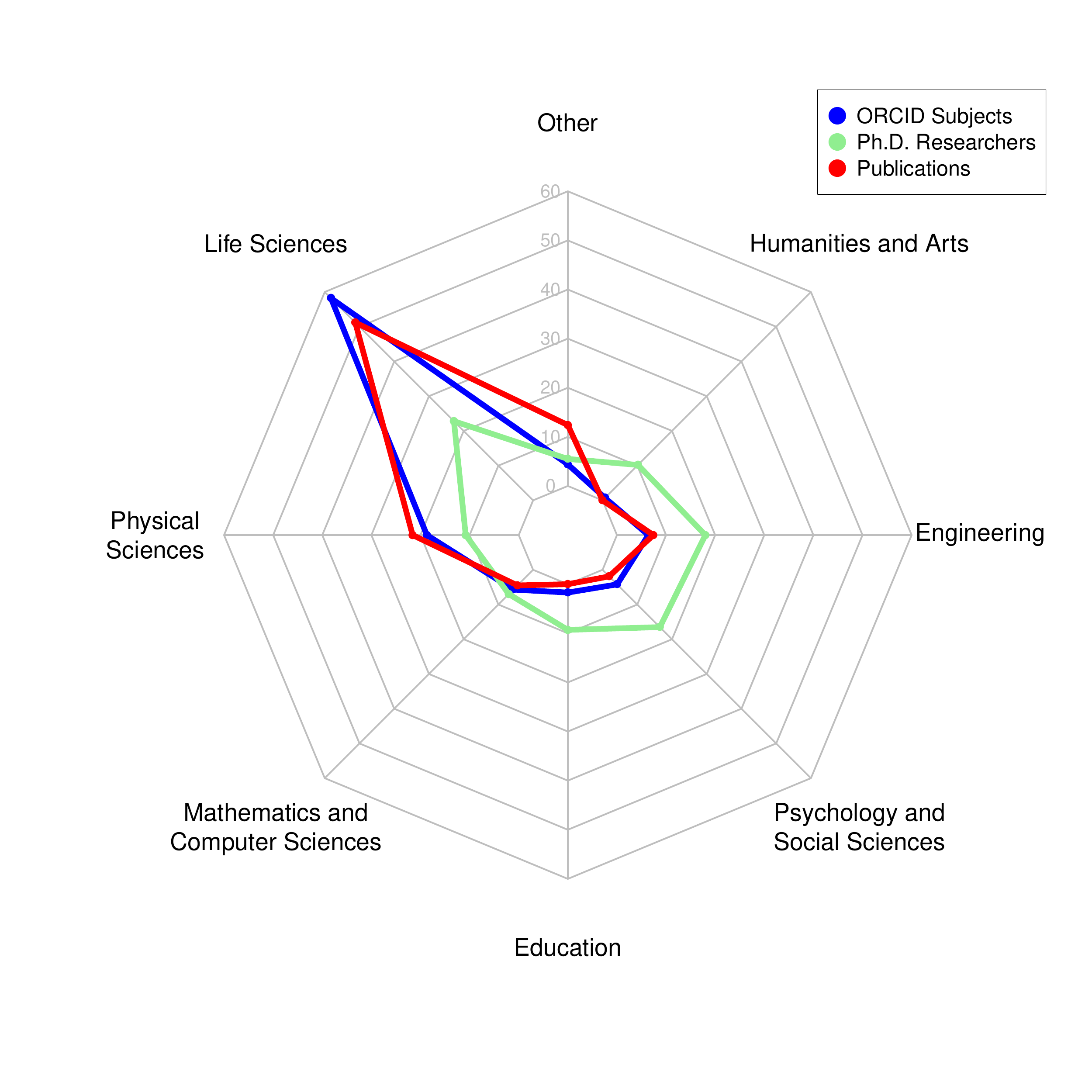}
\caption{Comparison of ORCID subjects based on the $2016$ dataset, fields of study of doctorate recipients 
in $2015$, and subjects of publications in the U.S. in $2014$}
\label{fig:sub_res_pub}
\end{figure}

Figure \ref{fig:sub_res_pub} shows the results of comparing the above data with subject data derived
from ORCID profiles from the $2016$ dataset.
The first thing that immediately becomes apparent is that the ORCID-specific data (in blue) and the 
UNESCO publication data (in red) are very similar. This seems to indicate that ORCIDs mirror the
scientific publication landscape fairly well.
In terms of specific subjects, we note that ``Life Sciences'' holds the top spot across all rankings. 
The percentage of doctoral degrees awarded, indicated in green, however, is less than half that 
of the ORCID-specific data and of the UNESCO publication data. Our interpretation of this finding is that 
there are proportionally many more life science researchers represented in ORCID than in the 
real world. 
We observe a similar pattern of over-representation of ORCID records for the area of ``Physical Sciences''
compared to the fraction of Ph.D. researchers.
On the other hand, the fields of ``Engineering'', ``Psychology and Social Sciences'', ``Education'', and 
``Humanities and Arts'' seems to be under-represented in ORCID records. The fraction of doctorate recipients 
in this area is much greater than the fraction of ORCID subjects.

It is important to note that Figure \ref{fig:sub_res_pub} conveys relative numbers. This means that even
though for a subject such as ``Mathematics and Computer Sciences'' the numbers are proportional, in terms
of absolute numbers, as shown in Section \ref{sec:cov_abs_numbers}, ORCID still needs to catch up.
\subsection{ORCID Subjects over Time}
The results from the previous section raise the question whether the ORCID subject distribution is
stable over time. If we saw significant movement in subject distribution over time, we could argue that
the subject coverage is likely to change in the future.
\begin{figure*}[h!]
\begin{subfigure}[b]{0.49\textwidth}
 \includegraphics[scale=0.38]{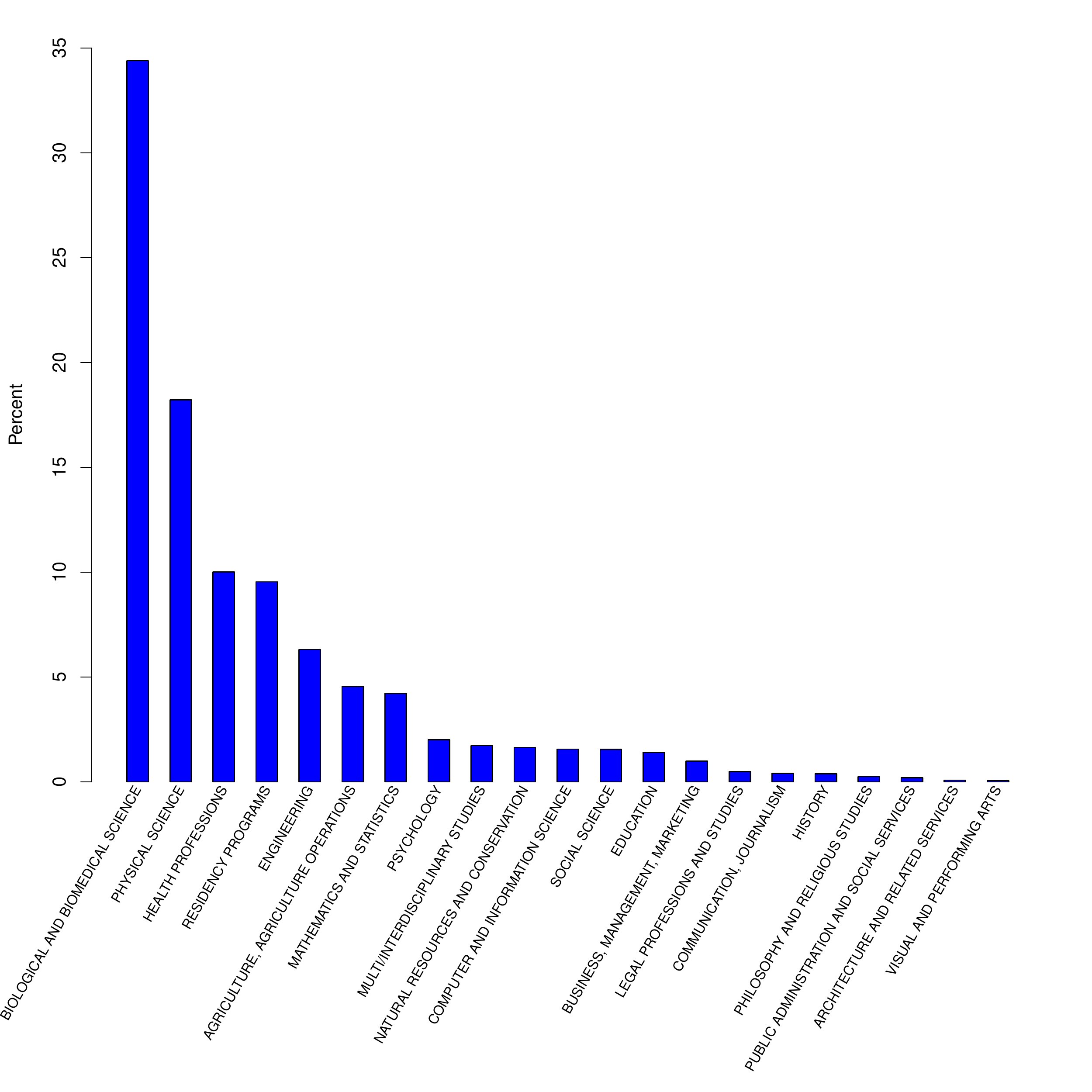}
 \caption{2013}
 \label{fig:sub_orcid_2013}
\end{subfigure}
\begin{subfigure}[b]{0.49\textwidth}
 \includegraphics[scale=0.38]{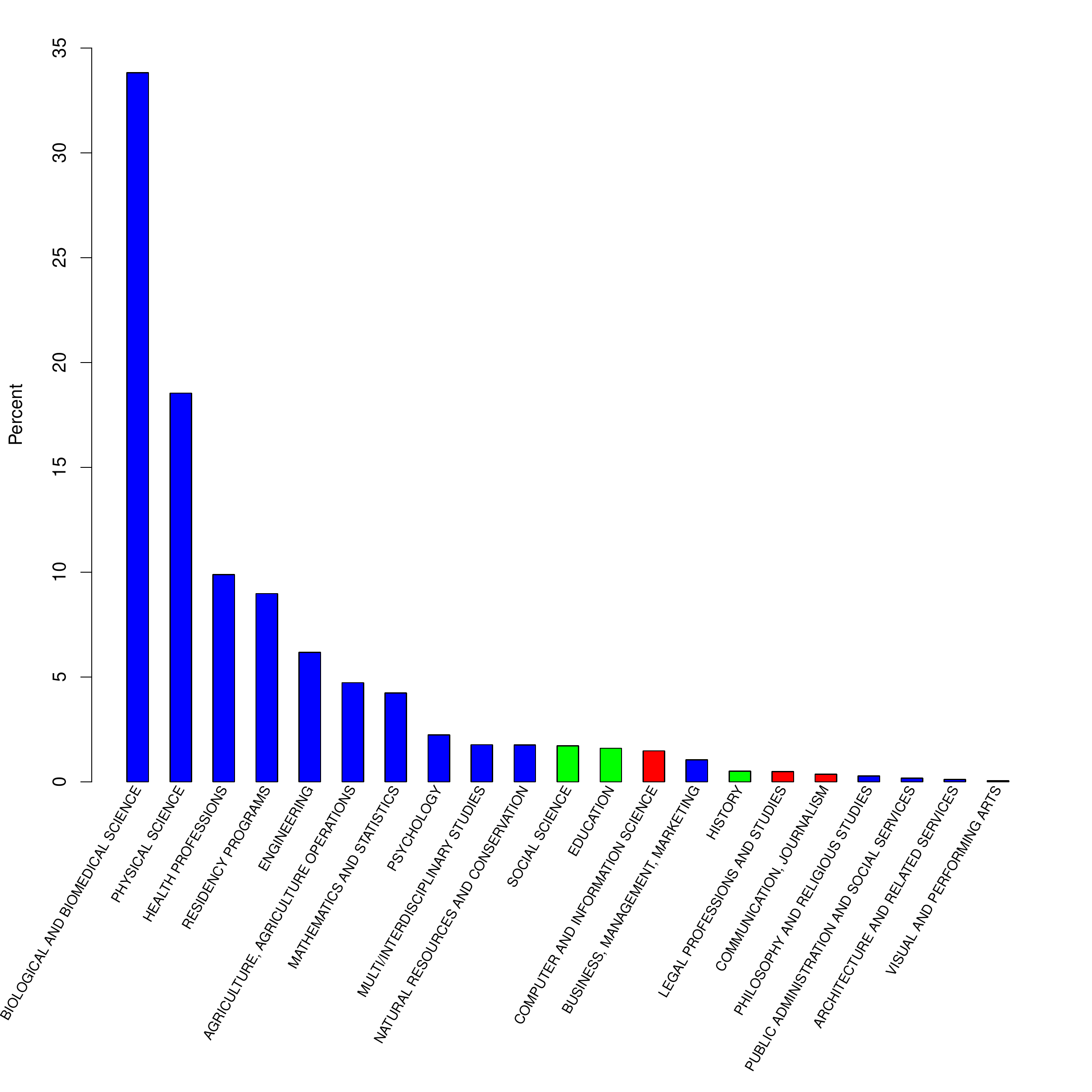}
 \caption{2014}
 \label{fig:sub_orcid_2014}
\end{subfigure}
\begin{subfigure}[b]{0.49\textwidth}
 \includegraphics[scale=0.38]{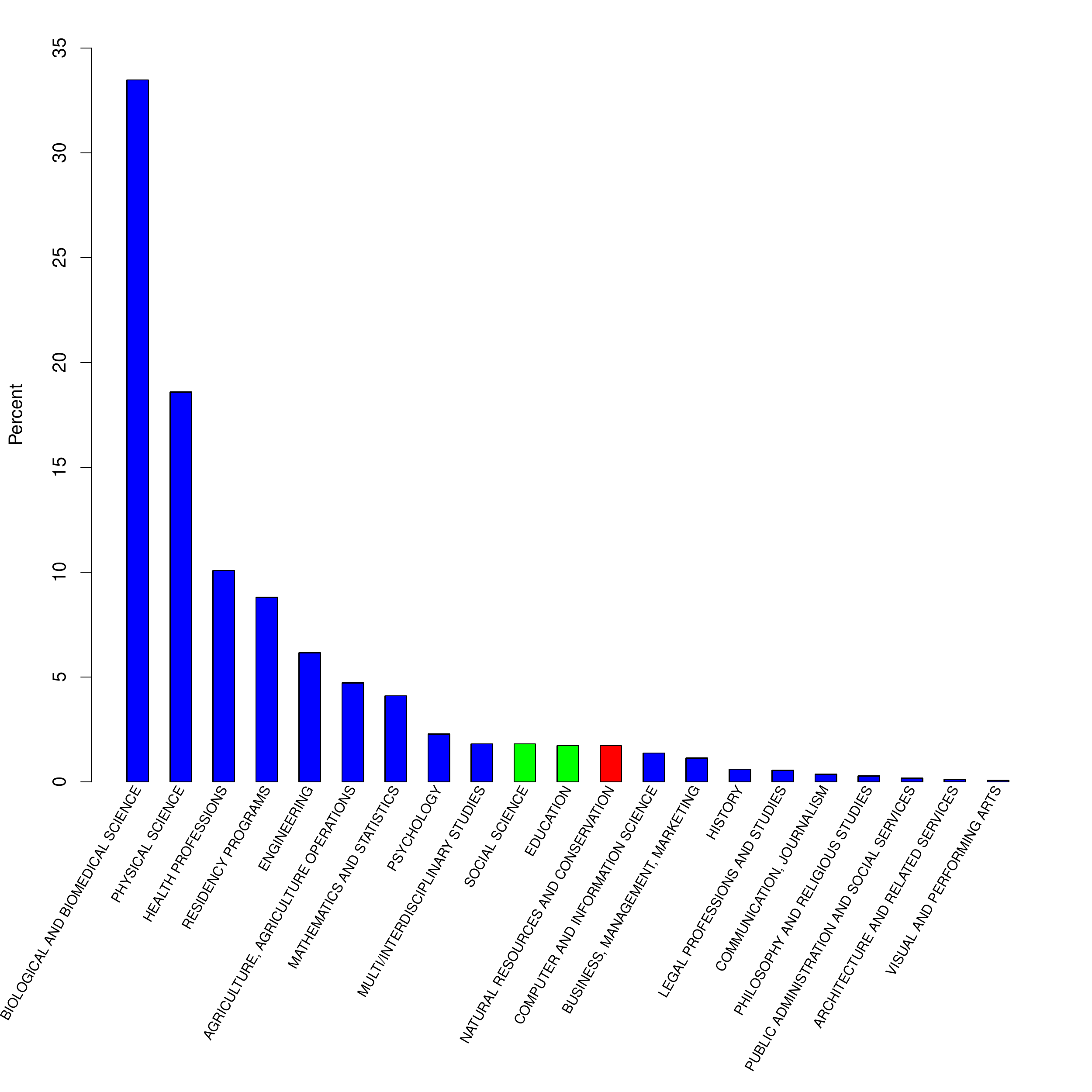}
 \caption{2015}
 \label{fig:sub_orcid_2015}
\end{subfigure}
\begin{subfigure}[b]{0.49\textwidth}
 \includegraphics[scale=0.38]{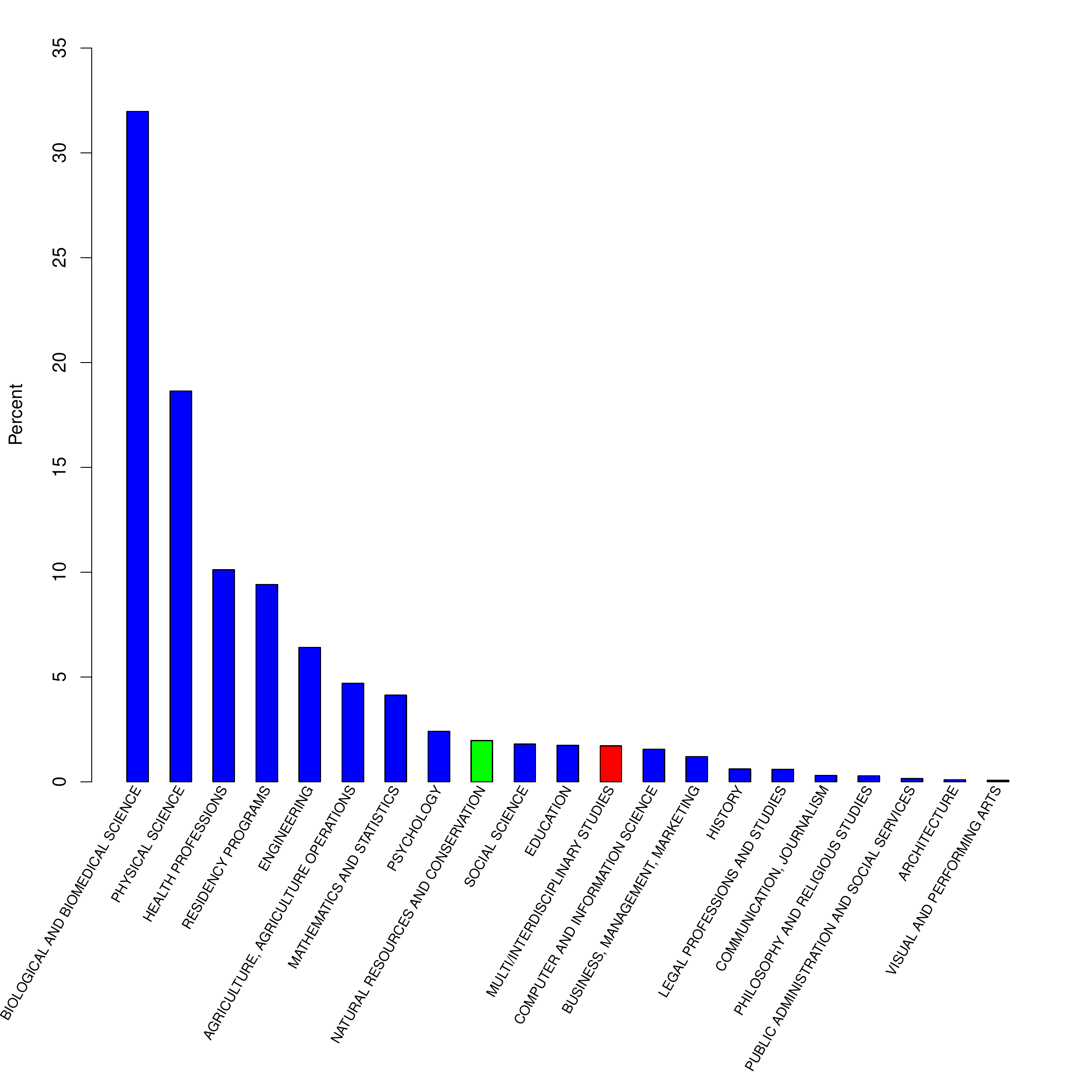}
 \caption{2016}
 \label{fig:sub_orcid_2016}
\end{subfigure}
\caption{Frequency of subjects represented in ORCID records and their changing ranks over time. Green 
bars represent a climb and red bars represent a drop in the ranking.}
\label{fig:sub_orcid_all_years}
\end{figure*}
Figure \ref{fig:sub_orcid_all_years} shows ORCID subject distributions for all four datasets.
From $2013$ (Figure \ref{fig:sub_orcid_2013}) on we can see a clear dominance of the medical fields. 
Three out of the top four subjects are from the medical area with ``Biological and Biomedical Science'' 
in the lead with around $35\%$. The subject ``Physical Science'' comes in second with $18\%$ followed 
by ``Health Professions and Related Programs'' and ``Residency Programs'' third and fourth with each 
around $10\%$. Together, the three medial fields make up for more than $53\%$ of all scores, which 
underlines their dominance. Other sciences such as engineering and mathematics get only around $5\%$ 
of the scores and other disciplines, for example, education, history and the performing arts get very 
few scores and therefore land at the tail end of the graph.

Figure \ref{fig:sub_orcid_2014} shows the subject ranking for the $2014$ dataset and also highlights the
changes in the ranking compared to the previous year. Subjects represented by blue bars have an unchanged 
rank compared to the previous year. Subjects with a green bar have climbed up the ranking and a red bar 
indicates a drop in the ranking. We see the top subjects mostly unchanged in both ranking and percentage 
of scores. Somewhat surprisingly, ``Social Science'', ``Education'', and ``History'' gained higher ranks 
whereas ``Computer and Information Science'' dropped.

Figures \ref{fig:sub_orcid_2015} and \ref{fig:sub_orcid_2016} show the distribution of subjects for the
$2015$ and $2016$ datasets, respectively. It is worth noting that ``Social Science'' and ``Education'' 
climbed yet again in the rankings in $2015$ and ``Natural Resources and Conservation'' jumped up the
ranking by three spots in $2016$.

All graphs in Figure \ref{fig:sub_orcid_all_years} confirm that ORCID records are dominated by the
medical field and physical sciences. They also show that there has been no change in the top subject 
ranks since the first available dataset in $2013$. Figure \ref{fig:sub_orcid_all_years} does not show
a lot of change in the subject distributions and hence does not indicate that an improved subject 
coverage can be expected in the near future.
%
%
%
\subsection{Geographical Coverage of ORCID Records}
In order to gain insight into the global coverage of researchers from a geographical point of view, we extract
the location information of the most recent affiliation per ORCID record. Table \ref{tab:orcid_geo} lists the 
top $20$ locations by country code. We can see that U.S. affiliations dominate the datasets with two European 
countries (Great Britain and Spain) being ranked second and third. The fact that China is only fourth ranked 
is surprising and indicates a much lower adoption rate there than elsewhere in the world. Brazil and India 
are following in the ranks.

The $2015$ UNESCO Science Report \cite{unesco2015} provides data on the world shares of researchers for selected 
countries in $2013$. The numbers are interesting as, for example, the ORCID representation for the U.S. ($16.9\%$
and $17.1\%$) is almost identical to the number reported by UNESCO ($16.7\%$). China, on the other hand
seems to be under-represented in the ORCID index where we only see $5.6\%$ compared to $19.1\%$ reported by UNESCO.
The same seems to hold true for Japan, Russia, and Germany.
The numbers for other countries in the report such as the United Kingdom ($3.3\%$), India ($2.7\%$), and Brazil 
($2.0\%$) are lower compared to what we find in the ORCID profiles. 
These results indicate that the geographical coverage of ORCID records does not fully mirror the worldwide picture.
In relative terms, the numbers for the U.S. are comparable but China and Japan are significantly under-represented. 
Other countries such as the United Kingdom, India, and Brazil appear to be over-represeted in ORCID.
\begin{table}
\caption{Geographical location distribution of affiliations in ORCID records}
\begin{tabular}{|c||c|c||c|c||c|c|} \hline
\textbf{Rank} & \multicolumn{2}{|c||}{\textbf{2015 ORCID}} & \multicolumn{2}{c||}{\textbf{2016 ORCID}} & \multicolumn{2}{c|}{\textbf{2013 UNESCO}} \\
& Geo & Freq & Geo & Freq & Geo & Freq \\ \hline \hline
1 & US & $16.9\%$ & US & $17.1\%$ & CN & $19.1\%$ \\ \hline 
2 & ES & $8.9\%$ & GB & $7.1\%$ & US & $16.7\%$ \\ \hline 
3 & GB & $6.4\%$ & ES & $6.9\%$ & JP & $8.5\%$ \\ \hline 
4 & CN & $5.6\%$ & CN & $5.6\%$ & RU & $5.7\%$ \\ \hline 
5 & IN & $5.5\%$ & BR & $5.6\%$ & DE & $4.6\%$ \\ \hline 
6 & BR & $4.4\%$ & IN & $5.2\%$ & KR & $4.1\%$ \\ \hline 
7 & PT & $4.3\%$ & IT & $4.1\%$ & FR & $3.4\%$ \\ \hline 
8 & IT & $4.3\%$ & AU & $3.1\%$ & GB & $3.3\%$ \\ \hline 
9 & AU & $2.8\%$ & PT & $3.0\%$ & IN & $2.7\%$ \\ \hline 
10 & SE & $2.6\%$ & RU & $2.9\%$ & CA & $2.1\%$ \\ \hline 
11 & RU & $2.4\%$ & SE & $2.0\%$ & BR & $2.0\%$ \\ \hline 
12 & KR & $2.1\%$ & UA & $2.0\%$ & TR & $1.1\%$ \\ \hline 
13 & DE & $1.9\%$ & DE & $2.0\%$ & IL & $0.8\%$ \\ \hline 
14 & UA & $1.8\%$ & KR & $1.9\%$ & IR & $0.8\%$ \\ \hline 
15 & CA & $1.8\%$ & CA & $1.8\%$ & AR & $0.7\%$ \\ \hline 
16 & FR & $1.7\%$ & FR & $1.8\%$ & MY & $0.7\%$ \\ \hline 
17 & JP & $1.5\%$ & IR & $1.7\%$ & EG & $0.6\%$ \\ \hline 
18 & IR & $1.4\%$ & JP & $1.5\%$ & MX & $0.6\%$ \\ \hline 
19 & TR & $1.4\%$ & TR & $1.4\%$ & ZA & $0.3\%$ \\ \hline 
20 & DK & $1.1\%$ & MX & $1.3\%$ & NA & NA \\ \hline 
\end{tabular}
\label{tab:orcid_geo}
\end{table}
\section{Richness of ORCID Profiles ($RQ2$)}
To address $RQ2$, we are now investigating the richness of ORCID profiles. For our paradigm,
profiles are rich when they contain web identities, further profile information about the scholar, 
as well as artifacts of potential interest to our efforts.
We examine web identities contained in ORCID profiles as they may lead to the discovery of in-scope 
artifacts in web portals where these identities were ultimatly minted. 
We further consider additional profile information that, using an algorithmic approach (see 
Section \ref{sec:intro}), may help facilitate the unveiling of web identities that in turn may again
help surface artifacts of interest.
Lastly, we analyze extracted artifacts as they may be orphans that are subject to archiving 
under our novel paradigm. 

ORCID records contain several metadata fields that are relevant for this investigtion. The values of
the fields ``Given Name'', ``Family Name'', and ``Affiliations'' (previously used for the geographical 
coverage assessment), can jointly be used to discover web identities with an algorithmic approach, 
as shown previously \cite{powell:egosystem}.
The field ``External URIs'' represents URIs that lead to web identities such as personal homepages,
a scholar's Twitter or LinkedIn page. 
The artifacts are extracted from the section in the ORCID profile called ``Works''. 

\begin{figure}
\includegraphics[scale=0.35]{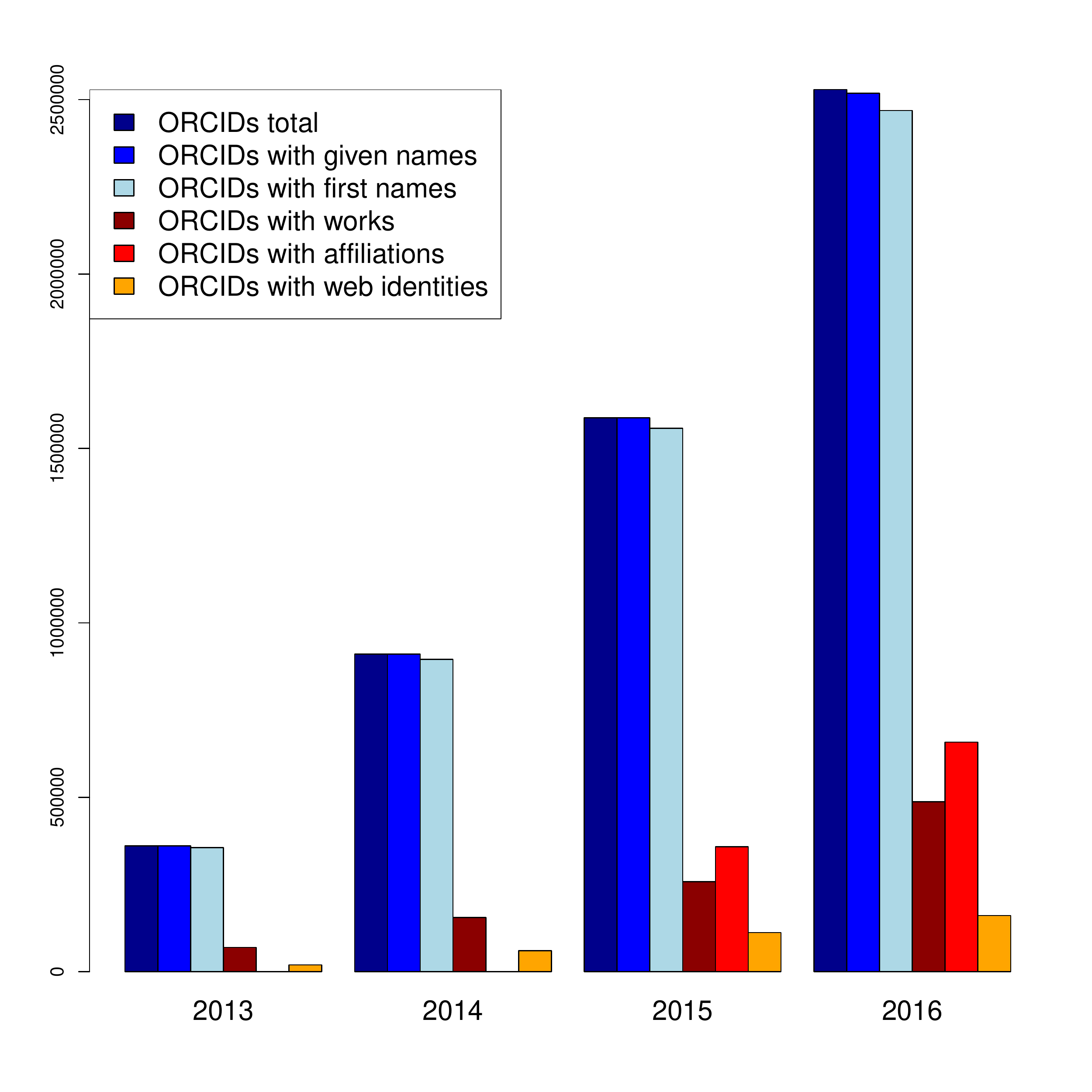}
\caption{Number of ORCID records with relevant information: given name, family name, works, affiliations, 
and web identities}
\label{fig:orcids_md_per_year}
\end{figure}

As a first step to evaluate the richness of ORCID profiles, we are interested in the number of ORCIDs
that actually contain the desired information. Figure \ref{fig:orcids_md_per_year} summarizes our findings.
Almost all ORCIDs contain a given and a family name but no affiliations are recorded in the $2013$ and $2014$
datasets. We notice a slow but steady increase of the number of ORCIDs with works ($19.3\%$ in $2016$),
affiliations ($26.0\%$ in $2016$), and web identities ($6.4\%$ in $2016$).
\subsection{Richness of Web Identities} \label{sec:rich_web_ids}
As seen in Figure \ref{fig:orcids_md_per_year}, a small percentage of ORCID records contain web identities.
Nevertheless, we are interested in extracting them and analyzing their type as they may lead to the discovery
of artifacts of interest. 
Each web entity in an ORCID profile has a type associated with it. Unfortunately, this type field is lacking
a controlled vocabulary, which makes this data very hard to interpret. Table \ref{tab:web_identity_labels}
lists the top $20$ web identity labels from the ORCID profiles of the $2016$ dataset. We immediately observe the
vocabulary problem as there are five different labels that describe presumably the same thing, a personal website
(Personal Website, Homepage, Home Page, Personal, Personal Webpage). The label issue aside, these references to 
personal websites are of interest to us as they potentially are artifact registries. Most likely, the majority of
them have a different structure so extracting information would require additional programmatic intelligence. 

Further, we recognize expected web identities such as LinkedIn, which is the most frequently found one and
Twitter. However, even these identities suffer from the vocabulary problem as, for example, ``LinkedIn'' and
``LinkedIn Profile'' make it into the list of the top $20$. We extracted other anticipated web identities 
such as SlideShare and FigShare but they are ranked $137th$ and $198th$, respectively, and hence did not make
it into Table \ref{tab:web_identity_labels}.

The fact that web identities are not particular common in ORCID profiles (see Figure \ref{fig:orcids_md_per_year})
combined with the label vocabulary problem for those that are available makes us conclude that the richness
of web identities required for our archival paradigm is not apparent.
It is worth noting though that the web identity labels may not be essential to extract and interpret web
identities. If an archival tool is aware of baseURIs of web portals, it could potentially match the identities
regardless of its label.
\begin{table}
\caption{Top 20 web identity labels from ORCID records in $2016$}
\begin{tabular}{|c|c|} \hline
\textbf{Label} & \textbf{Frequency} \\ \hline \hline
LinkedIn & 7326 ($3.8\%$) \\ \hline
Researchgate & 5306 ($2.7\%$) \\ \hline
Google Scholar & 4976 ($2.6\%$) \\ \hline
Personal Website & 4513 ($2.3\%$) \\ \hline
Homepage & 2916 ($1.5\%$) \\ \hline
Academia.edu & 2027 ($1.0\%$) \\ \hline
Research Gate & 1750 ($0.9\%$) \\ \hline
UCL IRIS Profile & 1657 ($0.9\%$) \\ \hline
Home Page & 1597 ($0.8\%$) \\ \hline
Twitter & 1487 ($0.8\%$) \\ \hline
LinkedIn Profile & 1433 ($0.7\%$) \\ \hline
KTH Profile & 1379 ($0.7\%$) \\ \hline
SISIUS & 1334 ($0.7\%$) \\ \hline
ID Dialnet & 1304 ($0.7\%$) \\ \hline
ID Personal SICA & 1286 ($0.7\%$) \\ \hline
Personal & 1261 ($0.6\%$) \\ \hline
Web de la Universidad & 1156 ($0.6\%$) \\ \hline
Personal Webpage & 1103 ($0.6\%$) \\ \hline
Blog & 1042 ($0.5\%$) \\ \hline
\multirow{2}{*}{Web de la Universidad} & \\
Polit\'{e}cnica de Madrid & 1012 ($0.5\%$) \\ \hline
\end{tabular}
\label{tab:web_identity_labels}
\end{table}
\subsection{Richness of Artifacts}
We are extracting information about artifacts from ORCID profiles by looking at
records of works. Figure \ref{fig:orcids_md_per_year} shows that a minority of ORCID records actually
contains information about a scholar's work, in fact, less than one in five ORCID records contain such
data. As a first result of this investigation, this does not imply a desired level of richness in ORCID
profiles.
%

Each work entry we do extract, however, contains a label that conveys the type of the work. This label
enables a high level disambiguation of the work and hence can help with the scoping of an artifact for our 
archiving paradigm. If the label, for example, conveys that a particular work is a publication of type 
``journal article'' we can, with some level of confidence, say that this work is out of scope for our 
approach as it stands a good chance to be convered by existing alternative archiving approaches such 
as LOCKSS, CLOCKSS, or Portico - approaches that are specialized in archiving journal articles.
\begin{table*}[h!]
\caption{ORCID work types over time}
\begin{tabular}{|c|c||c|c||c|c||c|c|} \hline
\multicolumn{2}{|c||}{\textbf{2013}} & \multicolumn{2}{c||}{\textbf{2014}} & \multicolumn{2}{c||}{\textbf{2015}} & \multicolumn{2}{c|}{\textbf{2016}} \\
Type & Freq & Type & Freq &Type & Freq & Type & Freq \\ \hline \hline
Journal Article & $93.4\%$ & Journal Article & $89.7\%$ & Journal Article & $86.7\%$ & Journal Article & $84.9\%$ \\ \hline 
Government Publication & $1.5\%$ & Conference Paper & $3.9\%$ & Conference Paper & $6.2\%$ & Conference Paper & $7.3\%$ \\ \hline 
Conference Proceedings & $1.4\%$ & Book & $1.7\%$ & Book & $2.3\%$ & Book & $2.5\%$ \\ \hline 
Chapter Anthology & $1.3\%$ & Book Chapter & $1.6\%$ & Book Chapter & $1.7\%$ & Book Chapter & $1.8\%$ \\ \hline 
Book & $0.9\%$ & Other & $1.0\%$ & Other & $1.1\%$ & Other & $1.4\%$ \\ \hline 
Other & $0.4\%$ & Standards And Policy & $0.6\%$ & Standards And Policy & $0.3\%$ & Conference Abstract & $0.3\%$ \\ \hline 
Patent & $0.1\%$ & Magazine Article & $0.2\%$ & Magazine Article & $0.2\%$ & Report & $0.2\%$ \\ \hline 
Manuscript & $0.1\%$ & Conference Poster & $0.2\%$ & Conference Abstract & $0.2\%$ & Magazine Article & $0.2\%$ \\ \hline 
Review & $0.1\%$ & Conference Abstract & $0.1\%$ & Conference Poster & $0.2\%$ & Conference Poster & $0.2\%$ \\ \hline 
Report & $0.1\%$ & Report & $0.1\%$ & Report & $0.1\%$ & Standards And Policy & $0.2\%$ \\ \hline 
\end{tabular}
\label{tab:orcid_work_types}
\end{table*}
Table \ref{tab:orcid_work_types} summarizes the top ten work types over time. The dominance of journal
articles is apparent for all four ORCID datasets. Conference papers as well as books and book chapters 
seem to be gaining in importance in the more recent past but still fade in comparison. 
It is important to note that the sort of artifacts that most likely would be in scope for our archiving 
paradigm are not well represented in ORCID records. For example, the type of work labeled ``Scholarly 
Project'' is ranked $20th$ in $2013$ and ``Artistic Performance'' is ranked $24th$ in $2016$.
The type ``Other'' may represent artifacts we are potentially interested in but since the label is very
ambiguous, these artifacts will need further evaluation.

With the rather low percentage of ORCID profiles containing works plus the fact that none of the top ranked 
work types are in scope for us, we realize that ORCID profiles lack the desired level of richness of artifacts.
We hence conjecture that, at this moment, the ORCID platform is not a good fit for step 2 in our high-level 
processes outlined in Figure \ref{fig:paradigm_concept}.
%
%
%
%
%
%
%
%
\section{Concluding Remarks}
We propose a novel archiving paradigm that is aimed at archiving web-based scholarly orphans.
The first and second step in this paradigm (Figure \ref{fig:paradigm_concept}) is focused on the 
discovery of web entities and artifacts in scope of our web archiving approach. 
Since ORCID has emerged as high potential scholarly web infrastructure that assigns web identities 
to scholars, allows listing additional web identities as well as artifacts per scholar, we were 
interested in determining whether it would be suitable as a discovery component in our archival 
processes.  
We approaches this work in two dimensions. First, we evaluated the coverage of ORCID in terms of 
number of researchers, in terms of subjects, and in terms of geographical coverage. Second, we 
analyzed the richness of ORCID profiles of information about a scholar, web identities, and 
artifacts.

We found that the ORCID subject coverage is proportional to subject coverage worldwide (as per 
publications) but in absolute numbers there is still significant room for growth.
We found more divergence with respect to the geographical coverage. Countries like China, Japan,
Russia, and Germany seem under-represented in ORCID.
%
However, we also discovered that ORCID growths at a very significant rate that outpaces the 
growth of researchers, for example. We therefore see a real chance that OCRID may achieve a 
level of coverage in the near future that is more suitable for our needs.

The results of the evaluation of the richness of ORCID profiles revealed that one out of 
five profiles contains information about the scholar's work. 
This number is surprisingly low and may indicate that scholars use other services such as 
ResearchGate or Academia.edu for their profile data. The majority of works we found in 
ORCID profiles are journal articles, which are out of scope for our use case.
Given these observations, it seems unreasonable to assume that researchers will eventually 
create entries for orphans in their profiles. The works component of ORCID profiles is
therefore less promising for our approach.
We further found that few profiles (less than $10\%$) contain web identities, which may be 
another indicator that researchers do not consider ORCID as their profile but rather as their 
identity. Nevertheless, since an ORCID is a web identity, it would make sense for ORCID to 
promote adding additional web identities so as to become an ``identity hub'' for researchers. 
This would be very beneficial for many use cases that involve access to machine-readable 
researcher profiles as it would allow to automatically navigate from a scholar's ORCID to 
their web presence in other portals.
To be able to better interpret web identities, however, it would help if a controlled vocabulary
for types could be used.
We acknowledge that ORCID profiles can provide rich data that can be used to algorithmically discover
other web identities of researchers. The given and family name(s) of scholars, their affiliations,
URIs to personal home pages (one of the most frequent web identities provided), and even subjects extracted
from their works can be used for this purpose. 

Clearly, ORCID adoption is on the rise but, at this point, relying on it as basic infrastructure 
for steps 1 and 2 in our paradigm is not an option. 
We are optimistic that the coverage will improve over time and eventually better align with researchers 
and research subjects.
To improve the richness of the profiles, some emphasis on promoting the addition of web identities would
be required from ORCID. We belive this aligns with ORCID's mission of umambiguously identifying researchers. 
%
%
\section{Acknowledgments}
This work is in part supported by the Andrew W. Mellon Foundation (grant number 11600663).
We would like to express our gratitude to the ORCID support staff, in particular Alainna Therese,
who provided invaluable feedback regarding components and history of ORCID records.
%
%
%

%
%
\end{document}